\documentstyle[11pt,titlepage]{article}

\newcommand{\sect}[1]{\setcounter{equation}{0}\section{#1}}

\newcommand{\be}{\begin{equation}}
\newcommand{\ee}{\end{equation}}

\newcommand{\bea}{\begin{eqnarray}}
\newcommand{\ena}{\end{eqnarray}}
\newcommand{\nn}{\nonumber}

\newcommand{\ba}{\begin{array}}
\newcommand{\ea}{\end{array}}

\newcommand{\bi}{{\bar{\imath}}}

\newcommand{\bk}{{\bar{k}}}
\newcommand{\bl}{{\bar{l}}}

\newcommand{\br}{{\bar{r}}}

\newcommand{\bA}{{\bar{A}}}

\newcommand{\bF}{{\bar{F}}}

\newcommand{\bM}{{\bar{M}}}

\newcommand{\bX}{{\bar{X}}}

\newcommand{\of}{\overline{F}}
\newcommand{\ovm}{{\overline{M}}}
\newcommand{\op}{{\overline{P}}}
\newcommand{\ow}{{\overline{W}}}
\newcommand{\ox}{{\overline{X}}}

\newcommand{\hhW}{\widehat{W}}

\newcommand{\what}{\widehat}

\newcommand{\bfr}{{\bf r}}
\newcommand{\bfs}{{\bf s}}
\newcommand{\bft}{{\bf t}}

\newcommand{\mT}{\mbox{{\bf T}}}

\newcommand{\ca}{{\cal A}}
\newcommand{\cb}{{\cal B}}
\newcommand{\cc}{{\cal C}}
\newcommand{\cd}{{\cal D}}

\newcommand{\cf}{{\cal F}}

\newcommand{\ck}{{\cal K}}
\newcommand{\cl}{{\cal L}}
\newcommand{\cm}{{\cal M}}

\newcommand{\car}{{\cal R}}

\newcommand{\cw}{{\cal W}}
\newcommand{\cy}{{\cal Y}}

\newcommand{\cab}{{\bar{\ca}}}
\newcommand{\cbb}{{\bar{\cb}}}
\newcommand{\ccb}{{\bar{\cc}}}
\newcommand{\cdb}{{\bar{\cal{D}}}}

\newcommand{\cwb}{{\bar{\cal{W}}}}

\newcommand{\ocf}{{\overline{\cf}}}
\newcommand{\ocw}{{\overline{\cw}}}

\newcommand{\al}{{\alpha}}
\newcommand{\bt}{{\beta}}
\newcommand{\gm}{{\gamma}}
\newcommand{\dt}{{\delta}}
\newcommand{\eps}{{\epsilon}}
\newcommand{\vep}{{\varepsilon}}
\newcommand{\la}{{\lambda}}
\newcommand{\si}{{\sigma}}
\newcommand{\th}{\theta}
\newcommand{\vp}{\varphi}
\newcommand{\om}{{\omega}}
\newcommand{\ze}{\zeta}

\newcommand{\dg}{{\dot{\gamma}}}
\newcommand{\da}{{\dot{\alpha}}}
\newcommand{\dv}{{\dot{\varphi}}}
\newcommand{\db}{{\dot{\beta}}}
\newcommand{\dd}{{\dot{\delta}}}

\newcommand{\Gm}{\Gamma}
\newcommand{\Si}{\Sigma}
\newcommand{\Dt}{\Delta}
\newcommand{\La}{\Lambda}

\newcommand{\Om}{{\Omega}}

\newcommand{\thb}{\bar{\theta}}
\newcommand{\etab}{\bar{\eta}}

\newcommand{\sib}{{\bar{\sigma}}}
\newcommand{\lab}{{\bar{\lambda}}}

\newcommand{\psib}{{\bar{\psi}}}
\newcommand{\chib}{{\bar{\chi}}}
\newcommand{\phib}{{\bar{\phi}}}
\newcommand{\zeb}{{\bar{\zeta}}}

\newcommand{\Psib}{{\bar{\Psi}}}

\newcommand{\Gmb}{\bar{\Gamma}}
\newcommand{\Sib}{\overline{\Sigma}}

\newcommand{\ogm}{{\overline{\Gm}}}
\newcommand{\ops}{{\overline{\Psi}}}

\newcommand{\undal}{{\underline{\alpha}}}
\newcommand{\undel}{\underline{\delta}}
\newcommand{\undbt}{{\underline{\beta}}}
\newcommand{\undgm}{{\underline{\gamma}}}

\newfont{\twelvemsb}{msbm10 scaled\magstep1}
\newfont{\eightmsb}{msbm8}
\newfont{\sixmsb}{msbm6}
\newfam\msbfam
\textfont\msbfam=\twelvemsb
\scriptfont\msbfam=\eightmsb
\scriptscriptfont\msbfam=\sixmsb
\catcode`\@=11
\def\Bbb{\ifmmode\let\next\Bbb@\else
  \def\next{\errmessage{Use \string\Bbb\space only in math mode}}\fi\next}
\def\Bbb@#1{{\Bbb@@{#1}}}
\def\Bbb@@#1{\fam\msbfam#1}
\newfont{\twelvegoth}{eufm10 scaled\magstep1}
\newfont{\tengoth}{eufm10}
\newfont{\eightgoth}{eufm8}
\newfam\gothfam
\textfont\gothfam=\twelvegoth
\scriptfont\gothfam=\eightgoth
\def\frak{\ifmmode\let\next\frak@\else
  \def\next{\errmessage{Use \string\frak\space only in math mode}}\fi\next}
\def\frak@#1{{\fam\gothfam{{#1}}}}
\catcode`@=12

\newcommand{\gom}{{\frak{M}}}

\newcommand{\rd}{{R^\dagger}}

\newcommand{\p}{{(\cd^{\al} \cd_{\al} - 8 \rd)}}
\newcommand{\q}{{(\cd_{\da} \cd^{\da} - 8 R)}}  

\newcommand{\ie}{{\em i.e. }}

\newcommand{\pp}{{(\cd^2 - 8 \rd)}}
\newcommand{\qq}{{(\cdb^2 - 8 R)}}  

\def\lp{\left(}
\def\rp{\right)}

\def\l[{\left[}
\def\r]{\right]}

\newcommand{\hs}{\hspace}

\newcommand{\cem}{\hspace{1cm}}

\newcommand{\ka}{{K\"{a}hler\ }}
\def\ym{\scriptscriptstyle{\cy \cm}}
\def\pot{\scriptscriptstyle{POT}}
\def\sm{\scriptscriptstyle{S+M}}
\def\m{\scriptscriptstyle{M}}

\newcommand{\f}[2]{{\textstyle\frac{#1}{#2}}}

\newcommand{\ci}[1]{\raise5pt\hbox{$\scriptstyle#1$}} 

\def\2#1{\mbox{#1}}

\def\12{\f{1}{2}}
\def\i2{\f{i}{2}}
\def\13{\f{1}{3}}
\def\s2{\sqrt{2}}

\newcommand{\prt}{\partial}

\newcommand{\ch}{\raise2.5pt\hbox{$\chi$}}
\newcommand{\chb}{\raise2.5pt\hbox{$\chib$}}

\newcommand{\vsp}[1]{\rule[- #1 mm]{0mm}{#1 mm}}

\newcommand{\xx}{Y}
\newcommand{\yy}{{\overline{Y}}}
\newcommand{\xxx}{y}
\newcommand{\yyy}{{\overline{y}}}

\begin{document}

\begin{titlepage}
\begin{center}
{\LARGE {\bf THE 3-FORM MULTIPLET IN SUPERGRAVITY}} 
\vskip 1cm

{\large P. Bin\'etruy, F. Pillon}\\
{\em LPTHE\footnote{Laboratoire associ\'e au CNRS-URA-D0063.},
Universit\'e Paris-XI, B\^atiment 211,  F-91405 Orsay Cedex, France} 
\\[.7cm] 

{\large G. Girardi} \\
\vsp{2}
{\em Laboratoire de Physique Th\'eorique}
{\small E}N{\large S}{\Large L}{\large A}P{\small P}\footnote{
URA 14-36 du CNRS, associ\'ee \`a l'E.N.S. de Lyon, et au L.A.P.P.} \\
{\em Chemin de Bellevue, BP 110, 
F-74941 Annecy-le-Vieux Cedex, France} \\[.7cm]

{\large R. Grimm} \\
{\em Centre de Physique Th\'eorique \\
C.N.R.S. Luminy, Case 907,
F-13288 Marseille Cedex 09, France} \\[1.2cm]      
\end{center} 

\centerline{ {\bf Abstract}}

\indent
We derive the couplings of the 3-form supermultiplet to the general
supergravity-matter-Yang-Mills system. Based on the methods of superspace
geometry,
 we identify component fields, establish their supergravity transformations and
construct invariant component field actions. Two specific applications
are adressed: the appearance of fundamental 3-forms in the context of
strong-weak duality and the use of the 3-form supermultiplets
to describe effective degrees of freedom relevant to the mechanism of gaugino 
condensation.

\vfill \rightline{LPTHE-Orsay \ \ 95/64}  
\rightline{{\small E}N{\large S}{\Large L}{\large A}P{\small
P}-A-553/95}  
\rightline{CPT-95/P.3258} 
\rightline{March 1996}

\end{titlepage}

\tableofcontents

\sect{INTRODUCTION}

The classical equivalence between a rank-two antisymmetric tensor and
a pseudoscalar is based on the Hodge duality between the field strength 
of a 2-form gauge field and the derivative of a 0-form.

A similar duality relation exists between a rank-3 antisymmetric
tensor and a constant scalar field. Such a relation was in particular
considered some time ago in connection with the cosmological constant
problem \cite{cosmo,DvN,Duff}.

The supersymmetric generalisation of this duality is particularly
interesting. In the first case, the rank-two antisymmetric tensor is
incorporated into the linear supermultiplet \cite{linear} which
includes also a real scalar and a Majorana  spinor field. The duality
transformation relates this supermultiplet to a chiral supermultiplet
\cite{duality} whose content includes the original scalar field as well as 
the pseudoscalar dual to the antisymmetric tensor.

Such a connection is often used in the context of superstring
theories. Indeed, the massless string modes include a dilaton and an
antisymmetric tensor which, together with a dilatino spinor field,
form a linear multiplet which plays an important role in the
effective field theory. A duality transformation is often performed
which turns these fields into a dilaton-axion (or rather a field with
axion-like couplings) system, central in all discussions of the
behavior of the theory under conformal and chiral transformations.

The role of supersymmetry is even more striking when one considers
the rank-three antisymmetric tensor. Whereas in the
non-supersymmetric case such a field does not correspond to any
physical degree of freedom (~through its equation of motion, its field
strength is a constant 4-form~), supersymmetry couples it with
propagating fields. Indeed, the 3-form supermultiplet \cite{1}
can be described by a chiral superfield $\xx$ and an antichiral field
$\yy$: 
\begin{equation}
\bar{D}^{\dot \alpha} \xx \ = \ 0, \cem D_{\alpha} \yy \ = \ 0,
\end{equation}
which are subject to the additional constraint:
\begin{equation}
D^2 \xx - {\bar D}^2 \yy \ = \ \f{8i}{3} \epsilon^{klmn} \Sigma_{klmn},
\label{eq:constraint}
\end{equation}
where $\Sigma_{klmn}$ is the gauge-invariant field strength of the
rank-three gauge potential superfield which we will denote by
$C_{klm}$: 
\begin{equation}
\Sigma_{klmn} \ = \ \partial_k C_{lmn} -  \partial_l C_{mnk} 
                  + \partial_m C_{nkl} -  \partial_n C_{klm}.
\end{equation}
Such a structure is obtained by solving the following constraint on the
super-4-form field strength:
\begin{equation}
\Sigma_{\underline \delta \underline \gamma \underline \beta A} \ = \ 0,
\label{eq:superconstraint}
\end{equation}
where $\underline \alpha$ is a (dotted or undotted) spinorial index
whereas $A$ is a  general superspace index. Such a constraint is
reminiscent of what is encountered in the case of supersymmetric
 Chern-Simons forms .

The component fields of the (anti)chiral
superfield $\xx$ and $\yy$ are propagating. Therefore, supersymmetry
couples the rank three antisymmetric tensor with {\em dynamical}
degrees of freedom, while respecting the gauge invariance associated
with the 3-form. 

Let us note that $\xx$ is not a general chiral superfield since it must
obey the constraint (\ref{eq:constraint}). Indeed, such a
constraint is possible only if $\xx$ derives from a prepotential
$\Om$ which is real:
\begin{equation}
\xx \ = \ -4 {\bar D}^2 \Om, \cem \yy \ = \ -4 D^2 \Om.
\label{eq:prepotential}
\end{equation}
All the preceding formulas find a generalisation in supergravity
theories \cite{2,3} which will  be the framework of the
present paper.

Rank-three antisymmetric tensors might play an important role in
several problems of interest, connected with string theories. One of
them is the breaking of supersymmetry through gaugino condensation.
As the formalism of the super-3-form is modelled along the lines of
the Yang-Mills Chern-Simons superforms, this should come as no surprise.
Indeed, in supersymmetric theories where the Yang-Mills fields are
coupled to a dilaton described by a linear multiplet -- such as
effective superstring theories --, the effective theory below the
scale of condensation is described by a chiral superfield subject
to the constraint (\ref{eq:constraint}); its scalar component is
the gaugino condensate itself. This chiral superfield derives from a
vector superfield such as in (\ref{eq:prepotential}), which is
interpreted (see the first reference  \cite{duality})
 as a ``fossile'' Chern-Simons superfield \cite{BGT}.

Another interesting appearance of the 3-form supermultiplet occurs
in the context of strong-weak coupling duality. More precisely, the
dual formulation of ten-dimensional supergravity \cite{Cham} appears
as an effective field theory of some  dual formulation of string
models, such as five-branes \cite{5brane}. The Yang-Mills field
strength which is a 7-form in ten dimensions may precisely yield in
four dimensions a 4-form field strength. The corresponding 3-form
may then play an important role in such a key issue as the cosmological
 constant problem \cite{cosmo}.

In the next section, we present the 3-form supermultiplet in the
context of supergravity. In particular, we give the explicit
solutions of the constraints (\ref{eq:superconstraint}) and present
the supersymmetry transformations. In section 3, we derive the
form of the action involving this supermultiplet coupled with
chiral supermultiplets. In section 4, we comment on
two types of applications where our analysis might apply: composite
3-form describing effective degrees of freedom below the gaugino
condensation scale and fundamental 3-form arising from the
compactification of the dual formulation of ten-dimensional
supergravity. 

\sect{THE 3-FORM AND SUPERGRAVITY}

\subsection{General definitions}

\indent

The superspace geometry of the 3-form multiplet has been
known for some time \cite{1}. Its coupling to the general
supergravity-matter system is most conveniently described
by generalizing the approach of \cite{1}
to the framework of $U_K(1)$ superspace \cite{2}. This means simply
that we have to deal with a 3-form gauge potential 
\be B^3 \ = \ \f{1}{3!} E^A E^B E^C B^3_{CBA}, \ee
where now $E^A$ denotes the frame of the full $U_K(1)$ superspace.
The 3-form gauge potential is subject to the gauge transformations
\be B^3 \ \mapsto \ {}^\La B^3 \ = \ B^3 + d \La, \ee
with parameters given as a superspace 2-form,
\be \La \ = \ \f12 E^A E^B \La_{BA}. \ee
The invariant field strength 
\be \Si \ = \ d B^3, \ee 
is a  4-form in superspace,
\be \Si \ = \ \f{1}{4!} E^A E^B E^C E^D \, \Si_{DCBA}, \ee
with coefficients
\bea 
\lefteqn{\f{1}{4!} E^A E^B E^C E^D \, \Si_{DCBA} \ = \ } \nn \\
&& \cem \f{1}{4!} E^A E^B E^C E^D \lp 4 \; \cd_D B^3_{CBA} + 6 \; T_{DC}{}^F
B^3_{FBA} \rp.
\ena
Here, the full $U_K(1)$ superspace covariant derivatives
and torsions appear. Likewise, the Bianchi identity,
\be d \Si \ = \ 0,\ee
is a $U_K(1)$ superspace five-form with coefficients 
\be 
\f{1}{5!} E^A E^B E^C E^D E^E 
\left( 5 \; \cd_E \Si_{DCBA} + 10 \; T_{ED}{}^F \Si_{FCBA} \right) \ = \ 0.
\ee
In these formulas we have kept the covariant differentials in
order to keep track of the graded tensorial structure of the
coefficients.

\subsection{Constraints and Bianchi identities}

\indent

The multiplet containing the  3-form gauge potential
is obtained after imposing constraints on the covariant
field-strength coefficients. Following \cite{1} we require
\be
\Si_{\undel \, \undgm \, \undbt \ A} \ = \ 0, \label{eq:sugraconst}
\ee
where $ \ \undal \sim \al, \da \ $ and $ \ A \sim a, \al, \da$.
The consequences of these constraints can be studied by
analyzing consecutively the Bianchi identities,
from lower to higher canonical dimensions
(\ie a spinor index contributes one-half while a vector
index contributes one in suitable units). The tensorial
structures of the coefficients
of $\Si$ at higher canonical dimensions are then
subject to restrictions due to the constraints.
In addition, covariant superfield conditions involving spinorial
derivatives will emerge. The contraints serve to reduce the number of 
independent component fields but do not imply any dynamical equations. 

As a result of this analysis, all the coefficients of the 4-form
field strength $\Si$ can be expressed in terms of the 
two superfields $\yy$ and
$\xx$, which are identified as follows in the tensorial decomposition:
\bea
\Si_{\dt \gm \ ba} & = & 
\f{1}{2} (\si_{ba} \eps)_{\dt \gm} \yy, \\
{\Si^{\dd \dg}}_{\ ba} & = & 
\f{1}{2} (\bar{\si}_{ba} \eps)^{\dd {\dg}} \xx.
\label{eq:7.12} 
\ena 
As we are working in $U_K(1)$ superspace these identifications
also allow to read off the {\em $U_K(1)$ weights} of $\xx$ and $\yy$, which are
\be \kappa (\xx) \ = \ +2, \cem \kappa(\yy) \ = \ -2, \ee
resulting in covariant (exterior) derivatives
\be D \xx \ = \ d \xx +2 A \xx, \cem D \yy \ = \ d \yy -2 A \yy, \ee
with $A = E^M A_M$ the $U_K(1)$ gauge potential. 
On the other hand, the {\em Weyl weights} are determined to be
\be \om(\xx) \ = \ \om (\yy) \ = \ +3. \ee
By a special choice of conventional constraints (\ie a covariant
redefinition of $B^3_{cba}$), it is possible to impose
\be
{{\Si_\dt}^{\dg}}_{\ ba} \ = \ 0.
\ee
The one spinor-three vector components of $\Si$ are given as
\bea
\Si{}_{\ \dt \ cba} &=& - \f{1}{16} \si^d_{\dt \dd} \ \vep_{dcba}
                          \cd^\dd \yy, \\
\Si^\dd{}_{\ cba} &=& \  +\f{1}{16} \bar{\si}^{d \dd \dt }
                          \, \vep_{dcba} \cd_\dt \xx.
\ena
At the same time one finds that the superfields $\yy$ and
$\xx$ are subject to the chirality conditions
\be
\cd_\al \yy \ = \ 0, \cem \cd^\da \xx \ = \ 0. 
\label{con1}
\ee
Moreover they are constrained by the relation
\be
\f{8i}{3} \vep^{dcba} \Si_{dcba} \ = \ 
      \left( \cd^\al \cd_\al - 24 \rd \right) \xx
    - \left( \cd_\da \cd^\da - 24 R \right) \yy. 
\label{con2}                 
\ee
This equation involving double spinorial derivatives is
a nontrivial restriction besides the chirality constraints,
because $\Si_{dcba}$ contains  (among other terms)
the curl of the purely vectorial coefficient of the 3-form.
As a consequence, its lowest superfield component is not an
independent field but is expressed in terms of other
components, as will be explained in detail in the next subsection. 

In conclusion, we have seen that all the coefficients of the superspace 
4-form $\Si$, subject to the constraints,  
are given in terms of the superfields  $\yy$ and $\xx$ 
and their spinorial derivatives. It is a matter of
straightforward computation to show that all the remaining Bianchi
identities do not contain any new information.

\subsection{Explicit solution of constraints : the unconstrained prepotential}

\indent

The analysis of the constraints via the Bianchi identities showed
how the 3-form superspace geometry is described in terms of the
superfields $\xx$ and $\yy$, themselves subject to the constraints
(\ref{con1}, \ref{con2}).
On the other hand, the direct solution of the constraints for the
3-form gauge potential allows to identify an unconstrained
prepotential. As a result, $\xx$ and $\yy$ given in terms of this 
unconstrained prepotential automatically satisfy 
(\ref{con1}, \ref{con2}).

We shall give here a brief sketch of this  and refer
to \cite{3} for a more detailed account of the  
explicit solution of the constraints in the case coupled to
supergravity. 

As an illustration of the method consider the constraints
\be \Si_{\dt \, \gm \, \bt \, A} \ = \ 0, \cem 
\Si^{\dd \, \dg \, \db}{}_A \ = \ 0, \ee
which are solved  in terms of prepotentials $U_{\bt A}$ and $V^\db{}_A$
such that, respectively,
\be
B^3_{\gm \bt A} \ = \ \cd_A U_{\gm \bt} + 
    \oint_{\gm \bt} \lp \cd_\gm U_{\bt A} + T_{A \gm}{}^F U_{F \bt} \rp,
\ee
and
\be
B^3{}^{\dg \db}{}_A \ = \ \cd_A V^{\dg \db} +
    \oint^{\dg \db} \lp \cd^\dg V^\db{}_A + T_A{}^{\dg \, F} V_F{}^\db \rp.
\ee
Since the prepotentials $U_{\bt A}$ and $V^\db{}_A$ should reproduce the gauge
transformations of the gauge potentials $B^3_{\gm \bt A}$ 
and $B^3{}^{\dg \db}{}_A$ they are assigned gauge transformations
\be U_{\bt A} \ \mapsto \ {}^\La U_{\bt A} \ = \ U_{\bt A} + \La_{\bt A}, 
\cem V^\db{}_A \ \mapsto \ {}^\La V^\db{}_A \ = \ V^\db{}_A +
\La^\db{}_A. \ee
On the other hand, the prepotentials are still allowed to change
under {\em pregauge transformations} which leave the gauge potentials
themselves unchanged. In the case at hand the pregauge transformations
are
\be U_{\bt A} \ \mapsto \ U_{\bt A} 
      + \cd_\bt \chi_A - (-)^a \cd_A \chi_\bt + T_{\bt A}{}^F \chi_F \ee
and
\be V^\db{}_A \ \mapsto \ V^\db{}_A 
      + \cd^\db \psi_A - (-)^a \cd_A \psi^\db + T^\db{}_A{}^F \psi_F. \ee
Playing around with these transformations it is quite straightforward
\cite{1} to convince oneself that the non-trivial unconstrained
prepotential $\Om$ is identified in
\be B^3{}_\gm{}^\db{}_a \ = \ -2i (\si_a \eps )_\gm{}^\db \Om, \ee
up to certain field dependent gauge transformations which
we have neglected here (for a more elaborate description see \cite{3}). 
Moreover, for the other non-vanishing components of the 3-form
gauge potential one finds
\be
B^3{}_{\gm \, ba} \ = \ 2 (\si_{ba})_\gm{}^\vp \, \cd_\vp \Om, \cem
B^3{}^\dg \, {}_{ba} \ = \ 2 (\sib_{ba})^\dg{}_\dv \, \cd^\dv \Om,
\ee
and, for the purely vectorial part,
\be
\lp [\cd_\al, \cd_\da] - 4 G_{\al \da} \rp \Om 
             \ = \ \f13 \si_{d\, \al \da} \vep^{dcba}
B^3{}_{cba}.
\ee             
Explicit substitution of these expressions for the 3-form
gauge potentials in the field strength gives  rise to
\bea \yy &=& -4 \p \Om, \\ \xx &=& -4 \q \Om. \ena
As already indicated above, these last three equations are
the explicit solution of the constraint equations
(\ref{con1},\ref{con2})\footnote{
Observe that for the special gauge choice $\Om=1$ one obtains
the identifications $\yy = 32 \rd$, $\xx=32 R$ and $-12 G^a
=\vep^{abcd}B^3{}_{cba}$.
In this case eq. (\ref{con2}) becomes simply
$\cd^2 R - \cdb^2 \rd = 4i \cd_a G^a$, one of the supergravity
equations (see for instance (B.88) in \cite{4}).}.
The explicit expressions for $\xx$ and $\yy$ illustrate also
the fact that
the prepotential remains undetermined up to a linear superfield,
\ie its pregauge-transformations
\be \Om \ \mapsto \ \Om' \ = \ \Om + \la, \ee 
are parametrized in terms of a linear superfield $\la$ which satisfies
\be \pp \la \ = \ 0, \cem \qq \la \ = \ 0. \ee

\subsection{Component fields and supergravity transformations}

\indent

We define the component fields as the lowest components of some superfields.
First of all, the three-index component field is identified as
\be B^3_{klm}|_{\th=\thb=0} \ = \ C_{klm}(x). \ee
As to the components of $\xx$ and $\yy$ we define
\be
\xx|_{\th=\thb=0} \ = \ \xx (x), \cem 
\cd_\al \xx|_{\th=\thb=0} \ = \ \s2 \, \eta_\al(x),
\ee
and
\be
\yy|_{\th=\thb=0} \ = \ \yy(x), \cem 
\cdb^\da \yy|_{\th=\thb=0} \ = \ \s2 \, \etab^\da (x). 
\ee
At the level of two covariant spinor derivatives we define
the component $H(x)$ as
\be
\cd^2 \xx|_{\th=\thb=0} + \cdb^2 \yy|_{\th=\thb=0} \ = \ -8 \,
H(x).
\label{defH}\ee
The orthogonal combination however is not an independent component field;
a look at our superspace geometry shows that it is given as 
\bea
\lefteqn{\cd^2 \xx|_{\th=\thb=0} - \cdb^2 \yy|_{\th=\thb=0} \ = \ 
    - \f{32i}{3} \vep^{klmn} \prt_k C_{lmn} } \nn \\[2mm]
&& \cem \cem + 2 \s2 i \, (\psib_m \sib^m)^\al \eta_\al 
   - 2 \s2 i \, (\psi_m \si^m)_\da \etab^\da \nn \\[2mm]
&& \cem \cem - 4 (\bM + \psib_m \sib^{mn} \psib_n) \, \xx
     + 4 (M + \psi_m \si^{mn} \psi_n) \, \yy.
\label{dC}
\ena
This expression illustrates also how the superspace approach
takes care of the modifications which arise from the 
coupling to supergravity, here the appearance of the Rarita-Schwinger
field and the supergravity auxiliary field, in the particular combination 
$M \yy - \bM \xx$ .

The component fields in the other sectors, \ie supergravity,
matter and Yang-Mills are defined as usual \cite{4}. 
Some new aspects arise in the treatment of the field dependent
$U_K(1)$ prepotential due to the presence of
the fields $\xx$ and $\yy$, carrying non-vanishing
$U_K(1)$ weights. It is for this reason that we refrain from
calling $K$ a \ka potential, we rather shall refer to
the field dependent  $U_K(1)$ prepotential as 
{\em kinetic prepotential}.

Before turning to the derivation of the
supergravity transformations we shall shortly digress
on the properties of the composite $U_K(1)$ connection
arising from the kinetic prepotential
\[ \ K(\phi,\xx,\phib,\yy) \]
subject to \ka transformations
\[ \ K(\phi,\xx,\phib,\yy) \ \mapsto \ K(\phi,\xx,\phib,\yy) + F(\phi) +
\bF(\phib). \]
Because of the non-zero $U_K(1)$ weight of the fields $\xx$ and $\yy$,
invariance of the \ka potential itself under $U_K(1)$ imposes the
condition
\be
\xx K_{\xx} = \yy K_{\yy}. \label{eq:Kinv}
\ee
We will make systematic use of this relation in what follows. An example
of a non-trivial \ka potential which satisfies this condition is
\be
K(\xx,\yy) = \ln ( 1 + \yy \xx), \label{eq:ex1}
\ee 
or, if we want to include some dependence on the matter fields $\phi$,
$\phib$
\be
K(\xx,\yy) = \ln \lp X(\phi,\phib) + Z(\phi,\phib) \, \yy \xx
\rp, \label{eq:ex2}
\ee
where $X$ and $Z$ are two functions of the matter fields.

Now recall first of all that the $U_K(1)$ connection 
component field is defined as the projection
to the lowest component of the superfield $A_m$,
\[ A_m(x) \ = \ A_m| \ = \ -\f12 \sib_m^{\da \al} A_{\al \da}| 
            + \f12 \psi_m{}^\al A_\al| + \f12 \psib_{m \, \da} A^\da |, \]
with
\[A_\al \ = \ \f14 E_\al{}^M \prt_M K, \cem 
  A^\da \ = \ - \f14 E^{\da \, M} \prt_M K, \]
and \[ A_{\al \da} \ = \ - \f{i}{8} [\cd_\al,\cd_\da] K + \f{3i}{2} G_{\al
\da}  . \]
Exploiting the (super)field dependence of the kinetic prepotential one finds
for the commutator term
\bea [\cd_\al,\cd_\da] K &=& 2i K_k \cd_{\al \da} \phi^k 
                            -2i K_\bk \cd_{\al \da} \phib^\bk 
                            +2i K_\xx \cd_{\al \da} \xx - 2i K_{\yy}
\cd_{\al \da} \yy \nn \\
&& +2 K_{\ca \cab} \cd_\al \Psi^\ca \cd_\da \Psib^\cab
   +6 \, (\xx K_\xx + \yy K_{\yy}) \,  G_{\al \da}, \ena
where $\Psi^\ca$ a short hand notation for $\phi^k, \xx$, 
and $\Psib^\cab$ for $\phib^\bk, \yy$. The important point is that
on the right hand the $U_K(1)$ connection appears in the covariant
derivatives of $\xx$ and $\yy$ due to the non-vanishing $U_K(1)$ weights.
Explicitly one has
\[ \cd_{\al \da} \yy|_{\th=\thb=0} \ = \ 
\si^m_{\al \da}\left( \prt_m \yy -2 A_m \yy - \f{1}{\s2} \psib_{m \, \da}
\etab^\da \right),  \]
\[ \cd_{\al \da} \xx|_{\th=\thb=0} \ = \ 
\si^m_{\al \da}\left( \prt_m \xx +2 A_m \xx - \f{1}{\s2} \psib_m{}^\al
\eta_\al \right). \]
Substituting in the defining equation for $A_m$ and factorizing gives then
rise to
\bea
\lefteqn{ A_m(x) + \f{i}{2} \, e_m{}^a b_a \ = \  
\frac{1}{4} \ \frac{1}{1-\xx K_\xx} 
\lp K_k \, \cd_m A^k - K_\bk \, \cd_m \bA^\bk \right. } \nn \\[1mm]
&& \left. + K_\xx \, \prt_m \xx - K_{\yy} \, \prt_m \yy  
+ i \, \sib_m^{\da \al} K_{\ca \cab} \, \Psi_\al^\ca \, \Psib_\da^\cab \ \rp .
\label{Am} \ena
Again, we have introduced a short hand notation: 
$\Psi_\al^\ca$ stands for $\chi_\al^k$ or $\eta_\al$, 
while $\Psib_\da^\cab$ stands for $\chib_\da^\bk$ or $\etab_\da$.
As is easily verified by an explicit calculation, $A_m$ defined this way
transforms as it should under the $U_K(1)$ transformations given above, \ie
\[ A_m \ \mapsto \ A_m + \f14 \prt_m (F-\bF).\]
Observe that the denominator accounts for the non-trivial
$U_K(1)$ phase transformations
\[\xx \ \mapsto \ \xx e^{-\f12(F-\bF)}, \cem 
        \yy \ \mapsto \ \yy e^{+\f12(F-\bF)} \]
of the 3-form scalar fields. In the following we shall frequently
use the particular combination
\be v_m(x) \ = \ A_m(x) + \f{i}{2} \, e_m{}^a b_a. \ee

We come now back to the issue of the supergravity transformations of the
component fields of the 3-form multiplet as defined above.
In general, in the spirit of \cite{5}, supergravity transformations
are defined as combinations of superspace diffeomorphisms 
(\ie Lie-derivatives in superspace as defined in some detail
in ref. \cite{6}) and field dependent
gauge transformations. In the case of the 3-form one has 
\be \delta B^3 \ = \ (\imath_\xi d + d \imath_\xi) B^3 + d \La
           \ = \ \imath_\xi \Si + d \left( \La + \imath_\xi B^3 \right), \ee
and the corresponding supergravity transformation is defined as a
diffeomorphism of parameter $\xi^A=\imath_\xi E^A$ 
together with a compensating
infinitesimal 2-form gauge transformation of parameter
\[\La \ = \ - \imath_\xi B^3, \]
giving rise to
\be \delta B^3 \ = \ \imath_\xi \Si \ = \ 
    \f{1}{3!} E^A E^B E^C \xi^D \Si_{DCBA}. \ee
The supergravity transformation of the component 3-form gauge
field $C_{klm}$ is then simply obtained from the double 
projection \cite{6} (simultaneously to lowest superfield
components and to space-time differential forms) as
\be \delta B^3 || \ = \ \f{1}{3!} dx^k dx^l dx^m \delta C_{mlk}
\ = \ \f{1}{3!} e^A e^B e^C \zeta^{\undel} \ \Si_{\undel \ CBA}|. \ee
Taking into account the definition $e^A = E^A||$ and the
particular form of the coefficients of $\Si$ one obtains
\be \delta C_{mlk} \ = \ \f{\s2}{16} 
\left( \zeb \sib^n \eta - \ze \si^n \etab \right) \vep_{nmlk} 
+ \f12 \oint_{mlk} \left[ (\psi_m \si_{lk} \ze) \, \yy 
                          +(\psib_m \sib_{lk} \zeb) \, \xx \right]. \ee

Let us turn now to the transformations of the remaining components.
To start, note that at the superfield level one has
\bea
\dt \xx &=& \imath_\xi d \xx \ = \ \imath_\xi D \xx -2 \imath_\xi A \, \xx, \\
\dt \yy &=& \imath_\xi d \yy \ = \ \imath_\xi D \yy +2 \imath_\xi A \, \yy.
\ena
Taking into account the explicit form of the field-dependent factor
$\imath_\xi A| = \ze^\undal A_\undal|$ one finds 
\bea 
\dt \xx &=& \s2 \ze^\al 
     \left\{ (1-\f12 \xx K_\xx)\eta_\al - \f12 \xx K_k \chi^k_\al \right\}
          +\f{1}{\s2} \zeb_\da \xx \nn \\
&&     \left\{ K_{\yy} \etab^\da + K_\bk \chib^{\da \bk}
\right\}, \\[2mm]
\dt \yy &=& \s2 \zeb_\da 
     \left\{ (1-\f12 \yy K_{\yy})\etab^\da -\f12 \yy K_\bk \chib^{\da
\bk}\right\} \nn \\
&&          + \f{1}{\s2} \ze^\al \yy 
     \left\{K_\xx \eta_\al + K_k \chi^k_\al \right\}
\ena
It is more convenient to use a notation where one keeps the
combination
\be \Xi \ = \  \ze^\undal A_\undal| \ = \ 
\f{1}{2\s2} \ze^\al \left( K_k \chi^k_\al + K_\xx \eta_\al \right) 
-\f{1}{2\s2} \zeb_\da \left( K_\bk \chib^{\da \bk} + K_{\yy} \etab^\da
\right), \ee
giving rise to a  compact form of the supersymmetry transformations
:
\be \dt \xx \ = \ \s2 \, \ze^\al \eta_\al -2 \, \Xi \, \xx, \cem 
    \dt \yy \ = \ \s2 \, \zeb_\da \etab^\da +2 \, \Xi \, \yy.  \ee
The transformation law for the 3-"forminos" comes out as\footnote{
          The covariant derivative $D_m$ differs from $\cd_m$
          in that it is given in terms of the particular combination 
          $v_m \ = \ A_m + \f{i}{2} e_m{}^a b_a$ \ie the covariant
          derivative $D_m$ does not contain the field $b_a$}
\bea
\dt \eta_\al &=& \s2 \ze_\al H 
                   +\f{4i\s2}{3} \ze_\al \vep^{klmn} \prt_k C_{lmn}
                   +i\s2 (\zeb\sib^m \eps)_\al D_m \xx 
                   - \Xi \, \eta_\al \nn \\
&&-\f{i}{2} \ze_\al \left( \psib_m \sib^m \eta - \psi_m \si^m \etab \right)
  -i \s2 (\zeb\sib^m \eps)_\al \psi_m{}^\vp \eta_\vp \nn \\
&& + \f{1}{\s2} \ze_\al \left\{ (\bM + \psib_m \sib^{mn} \psib_n) \, \xx
     - (M + \psi_m \si^{mn} \psi_n) \, \yy \right\},
\ena
and
\bea
\dt \etab^\da &=& \s2 \zeb^\da H +i \s2 (\ze \si^m \eps)^\da D_m \yy
                 -\f{4i\s2}{3} \zeb^\da \vep^{klmn} \prt_k C_{lmn} + \Xi \,
\etab^\da \nn \\
&& +\f{i}{2} \zeb^\da \left( \psib_m \sib^m \eta - \psi_m \si^m \etab \right)
    -i (\zeb\sib^m \eps)_\al \psib_{m \dv} \etab^\dv \nn \\
&& - \f{1}{\s2} \zeb^\da \left\{ (\bM + \psib_m \sib^{mn} \psib_n) \, \xx
     - (M + \psi_m \si^{mn} \psi_n) \, \yy \right\}.
\ena
Finally, the supergravity transformation of $H$ is given as
\bea
\dt H &=& \f{1}{\s2} (\zeb\sib^m)^\al D_m \eta_\al
          + \f12 (\zeb\sib^m \si^n \psib_m)
                   (D_n \xx - \f{1}{\s2} \psi_m{}^\vp \eta_\vp) \nn \\[1mm]
      & & \f{1}{\s2} (\ze \si^m)_\da D_m \etab^\da
          + \f12 (\ze \si^m \sib^n \psi_m)
                   (D_n \yy - \f{1}{\s2} \psib_{m \dv} \etab^\dv) \nn \\[1mm]
&& + \f{1}{3 \s2} \bM \ze^\al \eta_\al 
   + \f{1}{3 \s2} M \zeb_\da \etab^\da
   + \f{1}{3 \s2} (\zeb \sib^a \eta + \ze \si^a \etab) b_a \nn \\[1mm]
&& + \xx \, \zeb_\da \bX^\da|_{\th=\thb=0}  + \yy \, \ze^\al
X_\al|_{\th=\thb=0} 
   - \f{i}{\s2} (\zeb \sib^m \psi_m + \ze \si^m \psib_m) H \nn \\[1mm]
&& + \f{2}{3}(\zeb \sib^p \psi_p - \ze \si^p \psib_p)
                   \vep^{klmn} \prt_k C_{lmn} \nn \\[1mm]
&&  -\f{i}{4}(\zeb \sib^l \psi_l - \ze \si^l \psib_l)
           \left\{ (\bM + \psib_m \sib^{mn} \psib_n) \, \xx
         - (M + \psi_m \si^{mn} \psi_n) \, \yy \right\} \nn \\[1mm]
&&  -\f{1}{4 \s2}(\zeb \sib^n \psi_n - \ze \si^n \psib_n)
             (\psib_m \sib^m \eta - \psi_m \si^m \etab). 
\ena
   In this transformation law appear the lowest components of the superfields
$X_\al$ and $\bX^\da$.
These superfields which play a key role in the
construction of invariant actions, are defined as follows : 

\be X_\al \ = \ -\f{1}{8} \lp \cdb^2 - 8 R \rp \cd_\al K, \cem
\bX^\da \ = \ -\f{1}{8} \lp \cd^2 - 8 \rd \rp \cdb^\da K. \ee
One may now successively apply the spinorial derivatives to the kinetic
potential to evaluate the explicit form of these superfields.
Alternatively one may use the expression
\bea A &=& \f14 K_\ca D \Psi^\ca - \f14 K_\cab D \Psib^\cab 
            + \f{i}{8} E^a \sib_a^{\da \al} 
                  K_{\ca \cab} \cd_\al \Psi^\ca \cd_\da \Psib^\cab \nn \\ 
&&            + \f{3i}{2} E^a G_a \lp 1 - \f12 (\xx K_\xx + \yy K_{\yy})\rp, 
\ena
for the composite $U_K(1)$ connection, take the exterior derivative
$dA = F$ and identify $\bX^\da$ and $X_\al$ in the 2-form
coefficients
\be F_{\bt a} \ = \ + \f{i}{2} \si_{a \, \bt \db} \bX^\db 
                    + \f{3i}{2} \cd_\bt G_a, \cem
    F^\db {}_a \ = \ - \f{i}{2} \sib_a^{\bt \db} X_\bt 
                    + \f{3i}{2} \cd^\db G_a. \ee
A straightforward calculation then yields the component field
expression\footnote{
We make use, in the Yang-Mills sector, of the suggestive notations
\[
K_\bk \, (\lab^\da \! \! \cdot \! \bA)^\bk \ = \
      \lab^{\bfr \, \da} ( K_\bk \mT_\bfr{\, }^\bk{}_\bl \bA^\bl ), \cem  
K_k \, (\la_\al \! \cdot \! A)^k \ = \
       \la_\al^\bfr \lp K_k  \mT_\bfr{\, }^k{}_l A^l \rp. 
\]}
\bea
\lefteqn{\bX^\da (1-\yy K_{\yy})|_{\th=\thb=0} \ = \ 
     -\f{i}{\s2} K_{\ca \cab} \Psi_\al^\ca \sib^{m \, \da \al} 
            \lp D_m \Psib^\cab -\f{1}{\s2} \psib_{m \, \dv} \Psib^{\dv
\cab} \rp } \nn \\[1mm] 
&&   - \f{\s2}{8} \cd^2 \phi^k| \, K_{k \cab} \Psib^{\da \cab}  
   + \f{1}{\s2} H K_{\xx \cab} \Psib^{\da \cab}
   + \f{4i}{3 \s2} \Psib^{\da \cab} K_{\cab \xx} \vep^{klmn} \prt_k C_{lmn}
\nn \\[1mm]
&& -\f{1}{2 \s2} K_{\cab \cb \cc} \, \Psi^{\al \cc} \, \Psi_\al^\cb \,
\Psib^{\da \cab}
   - i K_\bk \, (\lab^\da \! \! \cdot \! \bA)^\bk \nn \\[1mm]
&&   - \f{i}{4} \Psib^{\da \cab} K_{\cab \xx} 
           (\psib_m \sib^m \eta - \psi_m \si^m \etab) \nn \\[1mm]
&&+ \f{1}{2 \s2} \Psib^{\da \cab} K_{\cab \xx}
         \left\{ (\bM + \psib_m \sib^{mn} \psib_n) \, \xx
         - (M + \psi_m \si^{mn} \psi_n) \, \yy \right\} 
\ena
and
\bea
\lefteqn{X_\al (1-\xx K_\xx)|_{\th=\thb=0} \ = \ 
         -\f{i}{\s2} K_{\ca \cab} \Psi^{\da \ca} \, \sib^m_{\al \da}  
            \lp D_m \Psi^\ca -\f{1}{\s2} \psi_m{}^\vp \Psi_\vp^\ca \rp} \nn
\\[1mm]
&&   - \f{\s2}{8} \cdb^2 \phib^\bk \, K_{\ca \bk} \Psi_\al^\ca   
   + \f{1}{\s2} H K_{\ca \yy  } \Psi_\al^\ca
   - \f{4i}{3 \s2} \Psi_\al^\ca K_{\ca \yy  } \vep^{klmn} \prt_k C_{lmn}
\nn \\[1mm]
&& -\f{1}{2 \s2} K_{\ca \cbb \ccb} \, \Psib_\da^\ccb \, 
                   \Psib^{\da \cbb} \, \Psi_\al^\ca
 + i K_k \, (\la_\al \! \! \cdot \! A)^k \nn \\[1mm]
&& + \f{i}{4} \Psi_\al^\ca K_{\ca \yy  }  
           (\psib_m \sib^m \eta - \psi_m \si^m \etab) \nn \\[1mm]
&&- \f{1}{2 \s2} \Psi_\al^\ca K_{\ca \yy  }
         \left\{ (\bM + \psib_m \sib^{mn} \psib_n) \, \xx
         - (M + \psi_m \si^{mn} \psi_n) \, \yy \right\}. 
\ena
 These are the component field expressions which are to be used in the
transformation 
law of $H$ (2.56). The same expressions wil be needed later on in the
construction 
of the invariant action.
\subsection{The matter $D$-term superfield}

\indent
For later convenience, we discuss now shortly the form of the $D$-term
superfield  $\cd^\al X_\al$ in the presence of the 3-form multiplet.
As pointed out in a previous subsection, the $U_K(1)$ gaugino
superfield $X_\al$ is given as
\bea
\lefteqn{2i \, X_\al \lp 1-\xx K_\xx \rp \ = \ } \nn \\[1mm]
& & K_{\ca \cab} \, \cdb^\da \Psib^\cab \, D_{\al \da} \Psi^\ca   
   - \f{i}{4} K_{\ca \cab} \, \cd_\al \Psi^\ca \, \cdb^2 \Psib^\cab \nn \\[1mm]
& & - \f{i}{4} K_{\ca \cbb \ccb} \, 
        \cdb_\da \Psib^\ccb \, \cdb^\da \Psib^\cbb \, \cd_\al \Psi^\ca
    -2i \, K_k \lp  \cw_\al \! \cdot \phi \rp^k.
\label{Xal} 
\ena
Note here that we are using the space-time covariant derivative $D_{\al
\da}$, which
by definition does not depend on the superfield $G_{\al \da}$. In full detail: 
\be \cd_{\al \da} \yy \ = \ D_{\al \da} \yy -3i \, G_{\al \da} \yy, \cem
   \cd_{\al \da} \xx \ = \ D_{\al \da} \xx +3i \, G_{\al \da} \xx, \ee
and
\bea
\cd_{\al \da} \, \cd_\db \yy &=& D_{\al \da} \, \cd_\db \yy 
   -\f{3i}{2} \, G_{\al \da} \, \cd_\db \yy, \\
\cd_{\al \da} \, \cd_\bt \xx &=& D_{\al \da} \, \cd_\bt \xx 
   +\f{3i}{2} \, G_{\al \da} \, \cd_\bt \xx.
\ena
In deriving the explicit
expression for $\cd^\al X_\al$ we shall make systematic use of this derivative,
which somewhat simplifies the calculations and will be useful anyway when
passing to
the component field expression later on. In applying the spinorial derivative
to (\ref{Xal}) it is convenient to make use of the following relations
\bea
\cd_\al \cdb_\da \yy &=& -2i \, D_{\al \da} \yy, \\
\cd_\al \cdb^2 \yy &=& -4i \, D_{\al \da} \cd^\da \yy 
                     +2 \, G_{\al \da} \, \cd^\da \yy -8 X_\al \yy, \\
\cd_\al \cdb^2 \phib^\bk &=& -4i \, D_{\al \da} \cd^\da \phib^\bk
      +2 \, G_{\al \da} \, \cd^\da \phib^\bk + 8 \lp  \cw_\al \! \cdot \phib
\rp^\bk.
\ena
In order to obtain a  compact form for $\cd^\al \! X_\al$, we
introduce $K^{\cab \ca}$ as the inverse of $K_{\ca \cab}$ and
we define
\bea
-4 \, F^\ca &=& \cd^\al \cd_\al \Psi^\ca 
                  + \Gm^\ca{}_{\cb \cc} \cd^\al \Psi^\cb \cd_\al \Psi^\cc, \\
-4 \, \bF^\cab &=& \cdb_\da \cdb^\al \Psib^\cab 
                  + \Gmb^\cab{}_{\cbb \ccb} \cdb_\da \Psib^\cbb \cdb^\da
\Psib^\ccb,  
\ena
with
\be
\Gm^\ca{}_{\cb \cc} \ = \ K^{\cab \ca} K_{\cab \cb \cc}, \cem
\Gmb^\cab{}_{\cbb \ccb} \ = \ K^{\cab \ca} K_{\ca \cbb \ccb}.
\ee
Moreover we define the new covariant derivatives
\bea
\nabla_{\al \da} \cd^\al \Psi^\ca &=& D_{\al \da} \cd^\al \Psi^\ca
         + \Gm^\ca{}_{\cb \cc} \, D_{\al \da} \Psi^\cb \, \cd^\al \Psi^\cc, \\
\nabla_{\al \da} \cdb^\da \Psib^\cab &=& D_{\al \da} \cdb^\da \Psib^\cab
         + \Gmb^\cab{}_{\cbb \ccb} \, D_{\al \da} \Psib^\cbb \, \cdb^\da
\Psib^\ccb.  
\ena
With these definitions the superfield expression of $\cd^\al \! X_\al$
becomes simply
\bea
\lefteqn{2i \, \cd^\al \! X_\al \lp 1-\yy K_{\yy} \rp \ = \ 
       4i \, \yy K_{\ca \yy  } \, X^\al \cd_\al \Psi^\ca 
      +4i \, \xx K_{\xx \cab} \, \bX_\da \cd^\da \Psib^\cab} \nn \\[1mm]
&& -2i \, K_{\ca \cab} \, D^{\al \da} \Psib^\cab \, D_{\al \da} \Psi^\ca
   - 4i \, K_{\ca \cab} \, F^\ca \bF^\cab \nn \\[1mm]
&&   -K_{\ca \cab} \,  \cdb^\da \Psib^\cab \, 
           \nabla_{\al \da} \cd^\al \Psi^\ca
     -K_{\ca \cab} \, \cd^\al \Psi^\ca \, 
           \nabla_{\al \da} \, \cdb^\da \Psib^\cab \nn \\[1mm]
&&  -\f{i}{4} \car_{\ca \cb \cab \cbb}
\, \cd^\al \Psi^\ca \, \cd_\al \Psi^\cb \, 
       \cdb_\da \Psib^\cab \, \cdb^\da \Psib^\cbb \nn \\[1mm]
&& -3i \, K_{\ca \cab} \, \cd^\al \Psi^\ca \, \cdb^\da \Psib^\cab \,
            G_{\al \da}\nn 
   +2i \, K_\bk \lp \cd^\al\cw_\al \! \cdot \phib \rp^\bk\\[1mm]
&& -4i \, K_{k \cab} \, \cdb_\da \Psib^\cab \, (\cwb^\da \! \cdot \phi)^k 
   -4i \, K_{\ca \bk} \, \cd^\al \Psi^\ca \, (\cw_\al \! \cdot \phib)^\bk.
\ena
This looks indeed very similar to the usual case. One of the
differences however is that the $F$-terms and their complex 
conjugates for the superfields $\xx$ and $\yy$ have special
forms. We will come back to this point later on in the discussion
of the full component field action. 

\subsection{The superpotential superfield}

The superpotential superfield is given as the combination
\be P \ = \ e^{K/2} \, W(\phi,\xxx), \ee 
with $\xxx$ defined as a holomorphic section
\be \xxx \ = \ e^{-K/2} \, \xx. 
\label{superpotential} \ee 
The superfield $P$ is covariantly chiral, $\ \cd^\da P=0 \ $,
and carries $U_K(1)$ weight $\kappa(P)=+2$.

We parametrize the covariant spinorial derivatives of $P$
such that
\be  \cd_\al P \ = \ \Si_\ca \, \cd_\al \Psi^\ca \ee
and
\be \cd^2 P \ = \ -4 \, \Si_\ca \, F^\ca 
                  + \Si_{\ca \cb} \, \cd^\al \Psi^\ca \, \cd_\al \Psi^\cb. \ee
The various components of the coefficients $\Si_\ca$
and $\Si_{\ca \cb}$ are given as 
\bea
\Si_{k} &=& e^{K/2} (W_k + K_k W) - \xx W_{\xxx} K_k
\label{eq:Sik}, \\
\Si_{\xx} &=& e^{K/2} W K_{\xx} + W_{\xxx} (1 - \xx K_{\xx}) 
\label{eq:SiX}
\ena
and
\bea
\Si_{kl} &=&  (e^{K/2} W - \xx W_\xxx)(K_{kl} + K_k K_l) \nn \\[1mm]
& & - \xx (W_{k \xxx} K_l + W_{l \xxx} K_k) 
    + e^{K/2} (W_{kl} + W_k K_l + W_l K_k) \nn \\[1mm]
& & + e^{-K/2} \, \xx^2 \, K_k K_l W_{\xxx \xxx} 
    -\Si_A {\Gamma^A}_{kl}, \label{eq:Sikl} \\[2mm]
\Si_{k \xx} &=&  (e^{K/2} W - \xx W_\xxx)(K_{k \xx} 
         + K_k K_\xx) \nn \\[1mm]  
& & + W_{k \xxx} \, (1-\xx K_\xx) + e^{K/2} W_k K_\xx \nn \\[1mm]
& & - e^{-K/2} \, \xx K_k W_{\xxx \xxx} \, (1-\xx K_\xx) 
    -\Si_A {\Gamma^A}_{k \xx},  \label{eq:SikX} \\[2mm] 
\Si_{\xx \xx} &=&  (e^{K/2} W 
       - \xx W_\xxx)(K_{\xx \xx} + K_\xx K_\xx) \nn \\[1mm]
& & + e^{-K/2} W_{\xxx \xxx} \, (1-\xx K_\xx)^2
    -\Si_A {\Gamma^A}_{\xx \xx}. \label{eq:SiXX}
\ena
Complex conjugate expressions are obtained from
\be
{\bar P} = e^{K/2} {\overline W}(\phib,\yyy),
\ee 
with $\yyy = e^{-K/2} \yy$.

\sect{THE COMPONENT FIELD ACTION}

\subsection{General action terms}

Our starting point for the construction of 
supersymmetric and $U_K(1)$ invariant component
field actions is the generic expression
\bea
\cl \ (\  r \ ,\ \br \ ) &=& 
- \f{1}{4} e \lp \cd^2 - 24 \rd \rp r|
- \f{1}{4} e \lp \cdb^2 - 24 R \rp \br| \nn \\[2mm]
& & + \f{i}{2} e (\psib_m \sib^m)^\al \ \cd_\al r|
    + \f{i}{2} e (\psi_m \si^m)_\da \ \cdb^\da \br| \nn \\[2mm]
& & - e (\psib_m \sib^{mn} \psib_n) \ r|
    - e (\psi_m \si^{mn} \psi_n) \ \br|,
 \label{genact}
\ena
with $r$ and $\br$ chiral resp. antichiral superfields
of $U_K(1)$ weights
\be \kappa(r) \ = \ +2, \cem \kappa(\br) \ = \ -2. \ee
Particular component field actions are then obtained
by chosing $r$ and $\br$ appropriately.
The complete action we are going to consider here will
consist of three separately supersymmetric pieces,
\be \cl_{total} \ \ = \ \ \cl_{\sm}
                    \ + \ \cl_{\pot}
                    \ + \ \cl_{\, \ym} \ , \ee
with
\be \cl_{\sm} \ = \ \cl \lp\ -3 R \ ,\ -3 \rd \ \rp, \ee
the combination of the supergravity action and the kinetic
terms of the matter sector, including the 3-form
multiplet,
\be \cl_{\pot} \ = \ 
         \cl \lp \ e^{K/2} W \ , \ e^{K/2} {\ow} \ \, \rp,\ee
the 3-form dependent superpotential and 
\be \cl_{\ym} \ = \ \cl \lp \ \f{1}{4} f_{\bf r s} \, 
          \cw^{{\bf r} \, \al} \, \cw^{\bf s}_\al \ ,  
           \ \f{1}{4} \bar{f}_{\bf r s} \, 
          \ocw^{\bf r}_\da \, \ocw^{{\bf s} \, \da} \ \rp, \ee
the Yang-Mills kinetic terms with 3-form independent gauge coupling
functions.
In the following we shall discuss one by one the three
individual contributions to the total Lagrangian.

The relevant superfield relations for the supergravity plus matter
kinetic actions are  
\be
-3 \, \cd_\al R \ = \ X_\al + 2 \, (T_{cb} \si^{cb} \eps)_\al, \hs{.5cm}
-3 \, \cd^\da \rd \ = \ \ox^\da + 2 \, (T_{cb} \sib^{cb} \eps)^\da
\ee
and
\bea
\lefteqn{\f{3}{4} \lp \cd^2 - 24 \rd \rp R
+ \f{3}{4} \lp \cdb^2 - 24 R \rp \rd \ = \  }\nn \\[2mm]
&& \hs{3cm}- \f12 R_{ba}{}^{ba} + 3 \, G^a G_a - 12 \, \rd R
- \f12 \cd^\al \! X_\al . \hs{1cm} 
\ena
It is then convenient to decompose the supergravity plus matter
action such that
\be
\cl_{\sm} \ = \ \cl_{supergravity} + e \, {\cal D_{\m}} \ ,
\ee
with the D-term matter component field ${\cal D_{\m}}$ defined as
\bea
{\cal D_{\m}} = -\frac{1}{2} \cd^\al \! X_\al | 
          + \frac{i}{2} \psi_m^\al \si^m_{\al \da} \ox^{\da} | 
          + \frac{i}{2} \psib_{m \da} \sib^{m \da \al} X_\al | \, .
\ena
The pure supergravity part is given by the usual expression, \ie
\bea
\lefteqn{\f{1}{e} \cl_{supergravity} \ = \ } \\
&& -\f12 \car 
    + \f12 \vep^{klmn} \lp \psib_k \sib_l D_m \psi_n
                           - \psi_k \si_l D_m \psib_n \rp
      -\f13 \ovm M +\f13 b^a b_a, \nn 
\ena
except that the $U_K(1)$ covariant derivatives 
of the Rarita-Schwinger field contain now 
the new composite $U_K(1)$ connection as defined above. For the
matter part one obtains
\bea
\lefteqn{(1 - \xx K_\xx) \, {\cal D_{\m}} \ = \ 
     - \s2 \, X^\al| \, \Psi^\ca_\al \ \yy K_{\ca \yy} 
     - \s2 \ \ox_\da| \, \ops^{\da \cab} \, \xx K_{\xx \cab} } \nn \\[1mm]
&&-g^{mn} K_{\ca \cab} \ D_m \Psi^\ca \ D_n \Psi^\cab
+ K_{\ca \cab} \ F^\ca \ \of^\cab \nn \\[1mm]
&& - \f{i}{2} \, K_{\ca \cab} \, \ops^{\da \cab} \ \si^m_{\al \da} 
       \nabla_m \Psi^{\al \ca}
- \f{i}{2} \, K_{\ca \cab} \, \Psi^{\al \ca} \ \si^m_{\al \da} 
       \nabla_m \ops^{\da \cab} \nn \\[1mm]
&& + \f14 \, \car_{\ca \cb \cab \cbb} \, \Psi^{\al \ca} \, \Psi^\cb_\al 
          \, \ops_\da^\cab \, \ops^{\da \cbb} 
   - \f12 \, K_{\ca \cab} \, \Psi^{\al \ca} \, \ops^{\da \cab} 
                   b_{\al \da} \nn \\[1mm] 
&& - \f{1}{\s2} \, (\psib_m \sib^n \si^m \ops^\cab) \  
        K_{\ca \cab} \ D_n \Psi^\ca 
- \f{1}{\s2} \, (\psi_m \si^n \sib^m \Psi^\ca) 
        \ K_{\ca \cab} \ D_n \ops^\cab \nn \\[1mm]
&& - (\psi_m \si^{mn}\Psi^\ca) \, K_{\ca \cab} \, (\psib_n \ops^\cab) 
   - (\psib_m \sib^{mn}\ops^\cab) \, K_{\ca \cab} \, (\psi_n \Psi^\ca) 
\nn \\[1mm]
&& - \f12 \, K_{\ca \cab} \, g^{mn} \, 
      (\psi_m \Psi^\ca) (\psib_n \ops^\cab) \nn \\[1mm]
&& - \lp {\bf D} \cdot \bA \rp^\bk K_\bk
   - i \s2 \, K_{k \cab} \, \ops_\da^\cab \, (\lab^\da \cdot A)^k
   + i \s2 \, K_{\ca \bk} \, \Psi^{\al \ca} \, (\la_\al \cdot \bA)^\bk
\nn \\[1mm]
&& - \f12 (\psib_m \sib^m)^\al \, K_k \, (\la_\al \cdot A)^k
   + \f12 (\psi_m \si^m)_\da \, K_\bk \, (\lab^\da \cdot \bA)^\bk, 
\ena
with the terms in the first line given as
\bea
\lefteqn{- \s2 \, X^\al| \, \Psi^\ca_\al \ \yy K_{\ca \yy} 
     - \s2 \ox_\da| \, \ops^{\da \cab} \, \xx K_{\xx \cab} \ = \ } \\[2mm]
 \frac{1}{1- \xx K_\xx}  \hs{-6mm} && \left[ 
  + i \, \yy K_{\cb \yy} \, K_{\ca \cab} \, \Psi^{\al \cb} \, \ops^{\da \cbb}
          \si^m_{\al \da} 
      \, \lp D_m \Psi^\ca - \f{1}{\s2} \, \psi_m{}^\vp \Psi^\ca_\vp \rp
     \right. \nn \\[1mm]
&& \left. \ + i \, \xx K_{\xx \cbb} \, K_{\ca \cab} \, \ops^{\da \cbb}
          \, \Psi^{\al \ca} \si^m_{\al \da} 
      \, \lp D_m \ops^\cab - \f{1}{\s2} \, \psib_{m \, \dv} 
                     \, \ops^{\da \cab} \rp  
         \right. \nn \\[1mm]
&& \left.
 - \xx K_{\xx \cbb} \, K_{\ca \cab} \, \ops_\da^\cbb 
        \, \ops^{\da \cab} F^\ca
 - \yy K_{\cb \yy} \, K_{\ca \cab} \, \Psi^{\al \cb} 
       \, \Psi^\ca_\al \, \of^\cab \right. \nn \\[1mm]
&& \left.\ - i \, \s2 \, \yy K_{\ca \yy} \, \Psi^{\al \cb} \, 
             K_k \, (\la_\al \cdot A)^k
            + i \, \s2 \, \xx K_{\xx \cab} \, \ops_\da^\cbb \,
              K_\bk \, (\lab^\da \cdot \bA)^\bk \ \right] \, . \nn 
\ena

Making use of the superpotential superfield and the corresponding
definitions given at the end of the previous section one derives
easily the component field expression
\bea
\f{1}{e}\cl_{\pot} &=&  \, \Si_\ca F^\ca 
        - \frac{1}{2} \, \Si_{\ca \cb} \Psi^{\al \ca} \Psi_\al^\cb 
        + \frac{i}{\s2} \, \Si_\ca \lp \psib_m \sib^m \Psi^\ca \rp \nn \\
&& -  \, e^{\frac{K}{2}} W \lp \bar{M} + \psib_m \sib^{m n} \psib_n \rp 
           \ \ + \ \ \mbox{h.c.}, \label{supercomp}
\ena
where $\Si_{\ca}$ and $\Si_{\ca \cb}$ are defined in 
(\ref{eq:Sik}-\ref{eq:SiXX}).

Finally the Yang-Mills component field action reads
\bea
\f{1}{e} \cl_{\, \ym} &=& -\f14 f_{\bfr \bfs} \left[ \frac{}{} 
      {\bf f}^{\bfr \, mn} \, {\bf f}^{\bfs}_{mn}
     + 2i \, \la^\bfr \si^m D_m \lab^\bfs
     + 2i \, \lab^\bfs \sib^m D_m \la^\bfr \right. \nn \\[1mm]
&& \hs{13mm} \left.
     - 2 \, {\bf D}^\bfr {\bf D}^\bfs + \f{i}{2} \vep^{klmn} \, 
             {\bf f}^{\bfr}_{kl} \, {\bf f}^{\bfs}_{mn}
   - 2 \, (\la^\bfr \si^a \lab^\bfs) \, b_a \frac{}{} \right] \nn \\[1mm]
&& -\f14 \frac{\prt f_{\bfr \bfs}}{\prt A^i} \left[ \frac{}{}
  \s2 \, (\ch^i \si^{mn} \la^\bfr) \, {\bf f}^{\bfs}_{mn} 
  - \s2 \, (\ch^i \la^\bfr) \, {\bf D}^\bfs 
  + (\la^\bfr \la^\bfs) \, F^i \right] \nn \\[1mm]
&& -\f14 \frac{\prt \bar{f}_{\bfr \bfs}}{\prt \bA^\bi} \left[ \frac{}{}
  \s2 \, (\chb^\bi \sib^{mn} \lab^\bfr) \, {\bf f}^{\bfs}_{mn} 
  - \s2 \, (\chb^\bi \lab^\bfr) \, {\bf D}^\bfs 
  + (\lab^\bfr \lab^\bfs) \, \of^\bi \right] \nn \\[1mm]
&& + \f18 \lp \frac{\prt^2 f_{\bfr \bfs}}{\prt A^k \prt A^l}
               -  \frac{\prt f_{\bfr \bfs}}{\prt A^i} \ \Gm^i{}_{kl} \rp
      (\ch^k \ch^l) (\la^\bfr \la^\bfs) \nn \\[1mm]
&& + \f18 \lp \frac{\prt^2 \bar{f}_{\bfr \bfs}}{\prt \bA^\bk \prt \bA^\bl}
          -  \frac{\prt \bar{f}_{\bfr \bfs}}{\prt \bA^\bi} 
             \ \ogm^\bi{}_{\bk \bl} \rp
      (\chb^\bk \chb^\bl) (\lab^\bfr \lab^\bfs) \nn \\[1mm]
&& \mbox{ plus $\psi_m, \, \psib_m$ dependent terms} ,
\ena
with the Yang-Mills field strength tensor
\be {\bf f}^\bfr_{mn} \ = \ 
    \prt_m a^\bfr_n - \prt_n a^\bfr_m 
        + a^\bfs_m \, a^\bft_n c_{\bfs \bft}{}^\bfr, \ee
and covariant derivatives of the gauginos
\bea
D_m \la^\bfr_\al &=& \prt_m \la^\bfr_\al 
            - \om_{m \, \al}{}^\vp \la^\bfr_\vp 
            + v_m \la^\bfr_\al 
            - a^\bft_m \la^\bfs_\al c_{\bfs \bft}{}^\bfr, \\[1mm]
D_m \lab^{\bfr \, \da} &=& \prt_m \lab^{\bfr \, \da}
            -\om_m{}^\da{}_\dv \lab^{\bfr \, \dv}
            -v_m \lab^{\bfr \, \da}
            -a^\bft_m \lab^{\bfs \, \da} c_{\bfs \bft}{}^\bfr.
\ena

\subsection{Solving for the auxiliary fields}

In the different pieces of the whole Lagrangian, we isolate
the contributions containing  auxiliary fields and proceed
sector by sector as much as possible.

 Diagonalization in $b_a$ makes use of the terms
\be 
\La_b \ = \ \f13 b^a b_a 
        -\f12 M_{\ca \cab} \lp \Psi^\ca \si^a \Psi^\cab \rp b_a
        +\f12 f_{\bfr \bfs} \lp \la^\bfr \si^a \lab^\bfs\rp b_a,
\ee 
with
\be M_{\ca \cab} \ = \frac{1}{1- \xx K_\xx} \ K_{\ca \cab}, \ee
whereas the relevant terms for the Yang-Mills auxiliary sector are
\bea
\lefteqn{\La_{\bf D} \ = \ \f12 f_{\bfr \bfs} {\bf D}^\bfr {\bf D}^\bfs
        + \frac{1}{1- \xx K_\xx} \,  
           {\bf D}^\bfs \lp K_\bl \, \mT_\bfs{\, }^\bl{}_\bk \, \bA^\bl \rp}  
\nn \\[1mm]
&& \hs{2cm}+ \f{\s2}{4} \, {\bf D}^\bfs 
     \left( \frac{\prt f_{\bfr \bfs}}{\prt A^k} (\ch^k \la^\bfr)
            +\frac{\prt \bar{f}_{\bfr \bfs}}{\prt \bA^\bk} 
                   (\chb^\bk \lab^\bfr) \right) .
\ena
The $F$-terms of chiral matter and the 3-form appear in the
general form
\be
\La_{F,\of} \ = \ F^\ca M_{\ca \cab} \of^\cab
         +F^\ca P_\ca + \op_\cab \of^\cab, 
\ee
with the definitions
\bea
P_k &=& \Si_k 
   -\frac{1}{4} \frac{\prt f_{\bfr \bfs}}{\prt A^k} (\la^\bfr \la^\bfs)
   - \xx M_{\xx \cbb} \, M_{k \cab} 
               \, \ops_\da^\cbb \, \ops^{\da \cab}, \\[1mm]
P_\xx &=& \Si_\xx - \xx M_{\xx \cbb} \, M_{\xx \cab} 
               \, \ops_\da^\cbb \, \ops^{\da \cab}.
\ena
We write this expression as
\be
\La_{F,\of} \ = \ \cf^k M_{k \bk} \, \ocf^\bk 
         - \op_{\! \cab} \, M^{\cab \ca} \, P_\ca
         + \cf^\xx \! \frac{1}{M^{\yy \xx}} \, \ocf^{\yy},
\label{quad} \ee
where $M^{\cab \ca}$ is the inverse of $M_{\ca \cab}$ and 
in particular
\be
 \frac{1}{M^{\yy \xx}}\ = \ M_{\xx \yy}-M_{\xx \bk} \,
 \gom^{\,\bk k} \,  M_{k \yy}, 
\ee
with $\gom^{\,\bk k}$ the inverse of $M_{k \bk}$. Moreover
\bea
\cf^k &=& F^k + \lp \op_\bk + F^\xx M_{\xx \bk} \rp \gom^{\,\bk k}, \\
\ocf^\bk &=& \of^\bk + \gom^{\,\bk k} \lp P_k + M_{k \yy} \of^{\yy} \rp,
\ena
and
\be
\cf^\xx \ = \ F^\xx + \op_\cab M^{\cab \xx}, \cem 
\ocf^{\yy} \ = \ \of^{\yy} + M^{\yy \ca} P_\ca.
\ee
We use now the particular structure of the 3-form
multiplet to further specify these $F$-terms.
Using (2.36), (2.37), (2.69), (2.70) we parametrise
\bea
\cf^\xx \ &=& \ H + i\lp \Dt + \frac{\ovm \xx - M \yy}{2i}
\rp + f^\xx,
 \\
\ocf^{\yy} \ &=& \ H - i\lp \Dt +\frac{\ovm \xx - M \yy}{2i
}\rp
 + \bar{f}^{\yy}
\ena
with
\bea
f^\xx &=& -\frac{1}{4}\Gm^\xx{}_{\cb \cc} \ \cd^\al \Psi^\cb \cd_\al \Psi^\cc
           + \op_\cab M^{\cab \xx}, \\
\bar{f}^{\yy} &=& -\frac{1}{4} \ogm^{\yy}{}_{\cbb \ccb} 
                    \ \cd_\da \ops^\cbb \cd^\da \ops^\ccb
           + M^{\yy \ca} P_\ca,
\ena
as well as
\bea
\Dt &=& \f{4}{3} \vep^{klmn} \prt_k C_{lmn}
- \f{1}{2 \s2} \lp \psib_m \sib^m \eta 
   - \psi_m \si^m \etab \rp \nn \\
&& + \frac{1}{2i} \left[ (\psib_m \sib^{mn} \psib_n) \, \xx
     - (\psi_m \si^{mn} \psi_n) \, \yy \right].
\ena
 
In terms of these notations the last term in (\ref{quad})
takes then the form
\bea
\cf^\xx \! \frac{1}{M^{\yy \xx}} \, \ocf^{\yy} \ &=& \
      \frac{1}{M^{\yy \xx}} \ \lp H+ \frac{ f^\xx + \bar{f}^{\yy}}{2} \rp ^2 
 \nn \\
    &+&   \ \frac{1}{M^{\yy \xx}}  
      \left(\Dt+ \frac{\ovm \xx - M \yy}{2i}
 +\frac{ f^\xx - \bar{f}^{\yy}}{2i}\right)^2 .
\ena
In this equation the last term makes a contribution to the
sector $M ,\ovm$ and the 3-form we consider next.
 Except for this term the sum of $ \La_b, \La_{\bf D}, \La_{F,\of} $ will give 
rise to the following diagonalised expression: 
\bea
\frac{1}{e} \cl (F^k, \of^\bk, b_a, {\bf D}^\bfr, H) &=&
 \frac{1}{3} \hat{b}_a \hat{b}^a + \frac{1}{2} \what{\bf D}^\bfr f_{\bfr \bfs}
\what{\bf D}^\bfs +\cf^k M_{k \bk} \, \ocf^\bk \nn \\
 &+&       \frac{1}{M^{\yy \xx}} \ \lp H+ \frac{ f^\xx + \bar{f}^{\yy}}{2}
\rp ^2 
- \frac{3}{16} \Bbb{B}_a \Bbb{B}^a 
\nn \\ &-&\frac{1}{2} 
\Bbb{D}_\bfr (f^{-1})^{\bfr \bfs} \Bbb{D}_\bfs 
 - \op_{\! \cab} \, M^{\cab \ca} \, P_\ca ,
\ena
 where $\hat{b}_a = b_a +\Bbb{B}_a $ with
\be
 {\Bbb{B}}^a =  - M_{\ca \cab} \lp \Psi^\ca \si^a \Psib^\cab \rp 
        + f_{\bfr \bfs} \lp \la^\bfr \si^a \lab^\bfs\rp ,
\ee
and ${\what{\bf D}}^\bfr = {\bf D}^\bfr +(f^{-1})^{\bfr \bfs} \Bbb{D}_\bfs
$ with
\bea
{\Bbb{D}}_{\bfr} &=& - \frac{1}{1- \xx K_\xx}   \lp K_k \,
{\mT}_{\bfr} .A \rp^k  
\nn \\
 &+& \f{\s2}{4} \, 
     \left( \frac{\prt f_{\bfr \bfs}}{\prt A^k} \; (\ch^k \la^\bfs)
            +\frac{\prt \bar{f}_{\bfr \bfs}}{\prt \bA^\bk} \; 
                   (\chb^\bk \lab^\bfs) \right) .
\ena       
Use of the equations of motion simply sets to zero the first
four terms leaving for the Lagrangian of (3.36)
\bea
\frac{1}{e} \cl  &=&
- \frac{3}{16} \Bbb{B}_a \Bbb{B}^a 
-\frac{1}{2} 
\Bbb{D}_\bfr (f^{-1})^{\bfr \bfs} \Bbb{D}_\bfs -{\op}_{\yy} \frac{1}{M_{\xx
\yy}} P_{\xx}
\nn \\
 &-& \lp \op_{\! \bk}  -  \op_{\!\yy} {\frac{M_{\xx \bk}}{M_{\xx \yy}}} 
 \rp 
 \, M^{\bk k} \, \lp P_k -{\frac{M_{k \yy}}{M_{\xx
\yy}}} P_{\xx} \rp,
\ena
where we have block diagonalised $ M^{\cab \ca}$. \\
As to the $M,\ovm$ dependent terms of the full action
we observe that they are intricately entangled with
the field strength tensor of the 3-form, a novel 
structure compared to the usual supergravity-matter
couplings. 
The relevant terms for this sector are identified to be
\bea
\La_{M,\ovm} &=&  3 e^K |W|^2 -\frac{1}{3} |M+3e^{K/2} W|^2
\nn \\ &+& \frac{1}{M^{\yy \xx}}\left[ \Dt -\frac{1}{2i} \lp
 M\yy-\ovm \xx \rp + \frac{1}{2i}\lp 
f^\xx - \bar{f}^{\yy} \rp  \right]^2 .\label{eq:LaMM}
\ena
\\
 One recognises in the first
two terms the usual superpotential contributions whereas the last
term is new and comes from (3.35).
This expression contains all the terms of the full action
which depend on $M$, $\ovm$ or the 3-form $C_{klm}$. 
The question we have to answer is as to how far
the $M$, $\ovm$  sector  and the 3-form
sector  can be disentangled, if at all.
Clearly, the dynamical consequences of this structure
deserve careful investigation.
\indent

The 3-form contribution is not algebraic, so one cannot
use the solution of its equation of motion ({\em e.o.m.})
 in the Lagrangian
\cite{Duff}. One way out is to derive the {\em e.o.m.}'s and
look for an equivalent Lagrangian giving rise to the same {\em
e.o.m.}'s. Explicitly  one obtains for the 3-form:
\be
 \prt_k  \left\{    \frac{1}{M^{\yy \xx}}\left[ \Dt -\frac{1}{2i} \lp
 M\yy-\ovm \xx \rp + \frac{1}{2i}\lp 
f^\xx - \bar{f}^{\yy} \rp  \right] \right\} =0 ,
\ee
solved by setting
\be
 \frac{1}{M^{\yy \xx}}\left[ \Dt -\frac{1}{2i} \lp
 M\yy-\ovm \xx \rp + \frac{1}{2i}\lp 
f^\xx - \bar{f}^{\yy} \rp  \right] =c \, , 
\ee 
where c is a real constant.
Then the {\em e.o.m.}'s for $M$ and $\ovm$ read
\be
M+3e^{K/2} W = -3ic \xx \cem ; \cem \ovm+3e^{K/2} \ow
=3ic \yy .
\ee
At last, one considers the {\em e.o.m.} for e.g. $\yy$, in
which we denote by $\cl(\yy)$ the many contributions of
$\yy$ to the Lagrangian, except for $ \La_{M,\ovm}$,
\be
\prt_m{\frac{\dt{\cl(\yy)}}{\dt{\prt}_m \yy}}
-\frac{\dt{\cl(\yy)}}{\dt \yy} -\frac{\dt  \La_{M,\ovm}}{\dt
\yy} =0 .
\ee
Using (3.42) and (3.43) the last term assumes the form
\bea
\frac{\dt  \La_{M,\ovm}}{\dt \yy}  &=& \frac{\dt }{\dt \yy} 
 \left\{ 3 e^K | W+icy|^2  -c^2 M^{\yy \xx} -ic( f^\xx -f^\yy)  \right. \nn \\
   && \cem \left.  - ic \left[ (\psib_m \sib^{mn} \psib_n) \, \xx
     - (\psi_m \si^{mn} \psi_n) \, \yy \right] \right\} .
\ena
This suggests that the equations of motion can be derived from an
equivalent Lagrangian
obtained by dropping the 3-form contribution and shifting the
superpotential $W$ to
 $W+icy$. This can be seen more clearly by restricting our attention to the
scalar
 degrees of freedom as in the next section.

\subsection{The scalar potential}

The analysis presented above allows to obtain the scalar potential of the
theory:
\bea
V &=&   \lp \Sib_{\! \bk}  -  (\Sib_{\!\yy} -ic) {\frac{M_{\xx \bk}}{M_{\xx
\yy}}} 
 \rp 
 \, M^{\bk k} \, \lp \Si_k \, -{\frac{M_{k \yy}}{M_{\xx
\yy}}} (\Si_{\xx} +ic) \rp 
\nn \\
 &+&({\Sib}_{\yy} -ic) \frac{1}{M_{\xx \yy}} (\Si_{\xx}+ic)
- 3 e^{K} |W+icy|^2
\nn \\
&+&\f12  \frac{1}{1- \xx K_\xx}    K_{\bk} \,
\lp {\mT}_\bfr .\bA \rp^{\bk}  (f^{-1})^{\bfr \bfs}  \frac{1}{1- \xx K_\xx}
   K_k \,
\lp{\mT}_\bfs .A \rp^k . \label{eq:V1}
\ena
We note that  the shift $W \mapsto W+icy $ induces  $\Si_k \mapsto \Si_k $ 
and $\Si_\xx \mapsto \Si_\xx +ic $, which are precisely the combinations
which appear in (\ref{eq:V1}). 

In fact (\ref{eq:V1}) is nothing but the scalar potential of some matter
fields 
$\phi^k$ of K\"ahler weight $0$ plus a field $\xx = y e^{K/2}$ of K\"ahler
weight $2$
with a superpotential $W+icy$ in the usual formulation of supergravity.
In order to show this, let us consider $\xxx$ and $\yyy$ as our new field
variables and define 
\be 
K(\xx, \phi, \yy, \phib) =\ck (\xxx, \phi, \yyy, \phib),
\ee
Taking as an example the \ka potential in (\ref{eq:ex1}), one 
finds\footnote{The new fields $\xxx$ and $\yyy$ are chiral when using
the derivatives covariant with respect to the new \ka potential
$\ck(\xxx,\yyy)$}
\bea
\xxx = {\xx \over (1 + \xx \yy)^{1/2}} &,&\;\;\; 
\yyy = {\yy \over (1 + \xx \yy)^{1/2}}, \nn \\\
\ck(\xxx,\yyy) &=& - \ln (1- \xxx\yyy). \label{eq:ex1'}
\ena
which is a typical \ka potential with $SU(1,1)$ noncompact symmetry.

We can express the matrix $M_{\ca \cab}$ and its inverse $M^{\cab \ca}$
in terms of the derivatives of $\ck$, namely $\ck_{\ca \cab}$ and of its inverse
$\ck^{\cab\ca}$ ( $\ca $ denotes $k,\xxx$ as well as $k, Y$
depending on the context). 
 Then it appears that the expression of the scalar
potential
becomes very simple as we use   the relevant 
relations collected in the appendix.
 Indeed, if we use the following definitions
\be
{\hhW} =W+ic\xxx, \cem \cem D_{\ca}{\hhW}= {\hhW}_{\ca}+ {\ck}_{\ca}{\hhW},
\ee
then
\bea V &=&e^{\ck} \lp D_{\cab}{\hhW} \, {\ck}^{\cab \ca} \,
D_{\ca}{\hhW}-3|\hhW|^2
\rp \nn \\
&+& \f12      {\ck}_{\bk} \,\lp {\mT}_\bfr .\bA \rp^{\bk}
  (f^{-1})^{\bfr \bfs}   {\ck}_k \,\lp{\mT}_\bfs .A \rp^k,
\ena
which is the familiar expression of the scalar potential of the scalar
fields $\phi_k$ and $y$ in the standard formulation of supergravity. 
\sect{APPLICATIONS}

\subsection{Fundamental 3-form}

\indent
  
Fundamental 3-forms naturally appear in the context of strong-weak
coupling duality. This can be seen most easily using the language of
five-branes \cite{5brane}. In the critical spacetime dimension $d=10$,
five-brane theories are conjectured to be dual to string theories in
the sense that a weakly coupled five-brane is a dual representation of
a strongly coupled heterotic string. After compactification to four
dimensions this may lead to a string/string duality.

The effective field theory corresponding to the five-brane scenario
would necessarily be described by the  formulation of supergravity
in $d=10$ dimensions with a seven-form field strength \cite{Cham}.
Under compactification, this would naturally yield a 4-form field
strength, {\em i.e.} the field strength of a fundamental 3-form.

In what follows, we will use a simple dimensional reduction
\cite{Witten,PB} to infer some of the couplings of this
fundamental 3-form as they arise from compactification. In $10$
dimensions the kinetic term for the six-form (dual of the 2-form
found among the massless string modes in the standard formulation)
involves the seven-form field strength we just referred to. We note
this field strength $K_{M_1 M_2 M_3 M_4 M_5 M_6 M_7}$ ($M_i = 1 \cdots
10$) and the corresponding action term reads \cite{Cham,5brane}: 
\be
S_{(10)} \ = \ \int d^{10} x \sqrt{-g^{(10)}} e^{\phi}
K^{M_1 M_2 M_3 M_4 M_5 M_6 M_7} K_{M_1 M_2 M_3 M_4 M_5 M_6 M_7},
\label{eq:10dlag}
\ee
where $\phi$ is the dilaton field and the upper indices are related
to the lower ones through the ten-dimensional metric tensor $g^{MN}$.

Under compactification to four dimensions, we recover our 4-form
field strength through the components:
\begin{equation}
\Sigma_{klmn} \equiv K_{klmnIJK},
\end{equation}
where the indices $I,J,K$ refer to the compact manifold. In our
simple compactification scheme \cite{Witten},
\begin{equation}
g_{IJ}^{(10)} = e^{\sigma} \delta_{IJ}, \cem
g_{mn}^{(10)} = e^{-3 \sigma} g_{mn},
\end{equation}
where $g_{mn}$ is the 4-dimensional metric and $\sigma$ is the
``breathing mode'' of the compact manifold.

The effective theory can be written in terms of the scalar fields
\begin{equation}
s = e^{-\phi / 2} e^{3 \sigma}, \cem t= e^{\phi / 2}
e^{\sigma},
\end{equation}
which are the dilaton and  modulus fields. In the present
formulation they are parts of respectively a chiral and a linear
supermultiplet \cite{PB}.The effective theory is described by  the
K\"ahler potential  \begin{equation} {\bf K} = - \ln s - 3 \ln t
\end{equation}
The action (\ref{eq:10dlag}) yields the following term in the
4-dimensional action 
\begin{equation}
S_{(4)} \sim \int d^4 x \sqrt{-g} st^3 \Sigma^{klmn} \Sigma_{klmn}
= \int d^4 x \sqrt{-g} e^{-{\bf K}} \Sigma^{klmn} \Sigma_{klmn}. 
\end{equation}
This should be compared with the corresponding term in (\ref{eq:LaMM}) (where
$\Delta$ contains a term proportional to $\Sigma^{klmn} \Sigma_{klmn}$).
It remains to be seen which field could be interpreted as the chiral
field $\xx$ appearing in the 3-form supermultiplet.

The field dependence of this kinetic term is given by $1/ M^{\yy\xx}$, whose
explicit form in terms of the fields $\phi^k$ and $\xxx$ is given by
(see appendix)
\be
1/ M^{\yy\xx} ={e^{-{\ck}} \over (1 + \xxx {\ck}_{\xxx})^2} 
{1 \over {\ck}^{\xxx \yyy} +\beta_k {\ck}^{\yyy k} +\beta_{\bk}
{\ck}^{\bk \xxx}+ \beta_k \beta_{\bk} {\ck}^{\bk k}},\label{eq:norm}
\ee
where $\beta_k$ and $\beta_{\bk}$ are defined in the appendix.
One recognises precisely the $e^{-\ck}$ dependence.

One might wonder which field of superstring models plays the role of the
field $\xxx$ accompanying the 3-form in the supergravity multiplet. In
compactification schemes such as Calabi-Yau manifolds where there is
only one independent 3-form, there should be a single $\xxx$ field. We
are working in the dual formulation of supergravity where \ka moduli are
in linear multiplets whereas the dilaton is in a chiral multiplet.
Therefore a natural candidate for $\xxx$ is the dilaton. Since the
dilaton \ka potential has a $SU(1,1)$ invariance, we would, under this
hypothesis, readily make the following identification:
\be
\xxx = {1-S \over 1+S}
\ee
with a \ka potential given by the example presented above for
illustrative purpose in (\ref{eq:ex1'}). One may worry that the terms
other than $e^{-\ck}$ in (\ref{eq:norm}) would induce an extra
dependence in $\xxx$ and thus in $S$. But with a \ka potential
(\ref{eq:ex1'}), one obtains 
\be
{1 \over (1 + \xxx {\ck}_{\xxx})^2} 
{1 \over {\ck}^{\xxx \yyy}}=1
\ee
Finally, \ka transformations $\xxx \rightarrow \xxx e^{-F}$
are related in this case to $SU(1,1)$ transformations on $S$ in a
straightforward manner:
\be
S \rightarrow { S cosh {F \over 2} + sinh {F \over 2} \over
S sinh {F \over 2} + cosh {F \over 2}}  .
\ee

It is also interesting to perform at this level a duality transformation in
order to see the content of the theory in the usual formulation.
This transformation reads at the level of the scalar fields:
\begin{equation}
st^3 \, \Sigma_{klmn} \sim \epsilon_{klmn} c,
\end{equation}
where $c$ is a scalar field (constant through its equation of
motion). The corresponding action term then reads:
\begin{equation}
{S_{(4)}}' = \int d^4 x \sqrt{-g} \, e^{\bf K} c^2.
\end{equation}
A term of this exact form was actually proposed in this context
\cite{DRSW} as a remnnant, in the 4-dimensional theory, of the field
strength of the 2-form: $H_{IJK}$ ($I,J,K$ compact indices). It is
known to break supersymmetry spontaneously \cite{Nilles}. 

\subsection{Composite 3-form: gaugino condensates.}

Another aspect of supersymmetry breaking where 3-forms play a role is
gaugino condensation. This is not completely surprising since the
constraints (\ref{eq:sugraconst}) on the 4-form field strength
$\Sigma_{ABCD}$ superfield in supergravity are imposed by analogy with
the case of a product of two Yang-Mills 2-forms $F_{AB} F_{CD}$. 
The corresponding 3-form is then the Chern-Simons form.

This appears most clearly in formulations of gaugino condensation
which involve a dilaton field, such as in superstring models: the
dilaton field is then incorporated into a linear multiplet $L$
\cite{BGT} (see also \cite{BDQQ}) in the fundamental theory. The
composite degrees of freedom are described, in the effective   theory
below the scale of condensation, by a vector superfield $V$ which 
incorporates also the components of the fundamental
linear multiplet $L$. The chiral superfield 
\be
U = - (\cd_\da \cd^\da - 8 R) V
\ee
has the same quantum numbers (in particular the same K\"ahler
weight) as the superfield $W^\alpha W_\alpha$. Its scalar component,
for instance, is interpreted as the gaugino condensate. 

Alternatively, the vector superfield is interpreted as a ``fossile''
Chern-Simons field which includes the fundamental degrees of
freedom of the dilaton supermultiplet. It can be considered as a
prepotential for the chiral superfield $U$: as such, its reality
imposes the constraint  (\ref{con2}) with $U=Y=\yy$.

{\ck} form of the superpotential is dictated by the anomaly structure
of the underlying theory \cite{VY} and is expressed as in 
(\ref{superpotential}) through the variable $u=U \exp (-K/2)$ which can
be understood in terms of the ratio of the infrared cut-off ($U^{1/3}$)
and the effective ultraviolet cut-off ($\exp (K/6)$). It reads simply
\cite{BGT}:
\be
W(u) = u \ln u
\ee
and its component form can be read off (\ref{supercomp}).

Certainly, the two applications just described deserve further study. 
In this paper,
we have restricted our attention to the derivation of the 
couplings of a 3-form supermultiplet to supergravity and we have tried
to be general enough in order to be able to describe the different
physical situations where such a supermultiplet might play a relevant
role,  somewhat neglected until now. 
\vskip 1cm
{\bf Acknowledgments}: P.B. wishes to thank the hospitality of the
Institute for Theoretical Physics
 (Santa Barbara) where part of this work was done.
\appendix
\section{\bf Appendix}

Here we gather some relations obtained as we use $y,\yyy$ as variables.
Let $K(\xx,\yy) = {\ck}(y,\yyy)$, with
\be
y=e^{-K/2} \yy, \cem \cem \xx =e^{{\ck}/2}y,
\ee
and the conjugate, then defining
\be
\alpha =\frac{1}{(1+y {\ck}_y)}, \cem \cem {\beta}_k =\frac{ y{\ck}_k}{ (1+y{\ck}_y)},
\ee
we obtain
\bea
 M_{\xx\yy} &=&e^{-{\ck}} \alpha^2 {\ck}_{y \yyy}, \nn \\
 M_{k \yy}  &=&e^{-{\ck}/2} \alpha( {\ck}_{k \yyy} -\beta_k {\ck}_{y \yyy} ), \nn\\
M_{k \bk}   &=& {\ck}_{k \bk} -\beta_k {\ck}_{y \bk} -\beta_{\bk} {\ck}_{k \yyy}
		+ \beta_k \beta_{\bk} {\ck}_{y \yyy}. 
\ena
This allows to compute the inverse of $M$ in terms of the inverse of ${\ck}$,
\bea
 M^{\yy\xx} &=&e^{{\ck}} \alpha^{-2} \lp {\ck}^{y \yyy} +\beta_k {\ck}^{\yyy k} +\beta_{\bk}
{\ck}^{\bk y}
		+ \beta_k \beta_{\bk} {\ck}^{\bk k} \rp, \nn \\
 M^{\yy k}  &=&e^{{\ck}/2}\alpha ^{-1} ( {\ck}^{ \yyy k} +\beta_{\bk} {\ck}^{\bk k} ), \nn \\
M^{k \bk}   &=& {\ck}^{ \bk k}.
\ena

The $\Sigma$'s appearing in the potential part also take simple forms
\be
{\Sigma}_{\xx} = \frac{D_y{\hhW}}{(1+y{\ck}_y)}, \cem \cem
{\Sigma}_k =D_k {\hhW} -\beta_k D_y {\hhW}.
\ee

\end{document}